# Plasmonic sensing with FIB-milled 2D micro-arrays of truncated gold nano-pyramids

Filippo Pisano[1,*], Antonio Balena[1,2], Muhammad Fayyaz Kashif[3], Marco Pisanello[1], Antonio Qualtieri[1], Leonardo Sileo[1], Tiziana Stomeo[1], Antonella D'Orazio[3,†], Massimo De Vittorio[1,2,†], Ferruccio Pisanello[1,†], Marco Grande[3,†]

[1]Fondazione Istituto Italiano di Tecnologia, Center for Biomolecular Nanotechnologies, Arnesano (LE), 73010, Italy
[2]Dipartimento di Ingegneria Dell'Innovazione, Università del Salento, Lecce, 73100, Italy
[3]Dipartimento di Ingegneria Elettrica e dell'Informazione, Politecnico di Bari, Bari, Italy
*Corresponding author: filippo.pisano@iit.it

†These authors jointly supervised this work

**Plasmonic platforms are a promising solution for the next generation of low-cost, integrated biomedical sensors. However, fabricating these nanostructures often requires lengthy and challenging fabrication processes. Here we exploit the peculiar features of Focused Ion Beam milling to obtain single-step patterning of a plasmonic two-dimensional (2D) array of truncated gold nano-pyramids (TNP), with gaps as small as 17 nm. We describe the formation of plasmonic bandgaps in the arrangement of crossed tapered grooves that separate the nano-pyramids, and we demonstrate refractive index sensing via dark field imaging in patterned areas of 30 μm × 30 μm.**

In the past few years, optical bio-sensing techniques based on surface plasmon resonances (SPR) have been widely applied in the detection of electro-chemical signals at extremely low molecular concentrations [1–4]. Among the many optical architectures that have been used for this purpose, periodic arrays of sub-wavelength apertures in continuous metal layers have emerged as a platform that supports a variety of resonant mechanisms [5–9]. In particular, the performances of SPR-based sensors have been improved using gratings structures such as gold nanotrenches [10] or gold nanoplatelets [11,12]. These nanostructures are commonly fabricated via Electron Beam Lithography (EBL) or Focused Ion Beam (FIB) milling. EBL is often preferred to FIB, as it offers a higher morphological quality, which in turn translates into a strong optical response [13]. FIB milling, instead, is more agile and better suited to pattern small areas on fragile substrates. This is because EBL requires many fabrication steps (resist deposition, resist exposure, resist development, metal deposition, lift-off), whereas FIB-based fabrication only needs two stages: metal deposition and milling [13]. On the other hand, FIB milling can introduce fabrication artifacts that interfere with the periodicity of the array and the repeatability of its single elements, due to lower resolution and to the re-deposition of milled material. One example is the tapered shape obtained when thin metals layers are patterned with $Ga^{2+}$ FIB systems, which unavoidably affect the obtained resonances [7,14]. Although this geometrical feature can be used to realize one-dimensional arrays of plasmonic nanotrenches, extending the approach to 2D plasmonic arrays represents a technological challenge, since the above-mentioned FIB limitations can result in low repeatability of the single element in the array.

In this work, we exploit the peculiar tapered shape of FIB-milled grooves to obtain a plasmonic 2D array of truncated gold nano-pyramids (TNP) in a single patterning step (Fig. 1a). The fabrication method can achieve nanogaps as narrow as 17 nm at the pyramids basement with $Ga^{2+}$ FIB, and we model the formation of plasmonic bandgaps in the 2D arrangement of sharp tapered grooves that results from FIB milling (Fig. 1b) [7,14], finding that the V-shape of the pattern enhances the structures' response. We then confirm our prediction experimentally and we show that a small 30 μm × 30 μm TNP sensor enables the detection of refractive index (RI) changes via dark field imaging. This, in turn, demonstrates that our approach is readily scalable to produce high-density arrays of independently addressable bio-sensing elements.

TNP arrays were fabricated in a gold film deposited on a borosilicate glass microscope coverslip (Fig. 1a). The process involved two steps. First, a 5-nm thick adhesion layer of Cr and a 180 nm-thick layer of Au (T) were deposited on a thin glass substrate via thermal deposition. In accordance with McPeak et al. [15], the deposition was performed at a rate of 1 Å/s, with chamber pressure of $6×10^{-6}$ mbar. Second, the nano-pyramids were carved in the gold layer by milling thin slits of metal with a $Ga^{2+}$ FIB (FEI dual-beam HeliosNanoLab600i) system. The ion beam current was set at

7.7 pA with 30 kV voltage. The beam spot (12 nm in diameter) was scanned sequentially in the two milling directions, covering one gap at a time across the full array length, with minimal line overlap (5-10 %). All vertical (V) slits were milled before fabricating horizontal (H) slits. The fabrication of a single array, 30 μm×30 μm wide, lasted for approximately 30 minutes. The morphological characterization of the final device was verified through Scanning Electron Microscope (SEM) inspection. As shown in Fig.1c, this process produced arrays with gaps of width g∼100 nm at the gold top surface. The platelets' shape was influenced by the milling protocol. In fact, the H direction, milled last, had slightly larger gap widths than the V direction. In addition, platelets boundaries in the V direction, milled first, had a tendency to assume a convex shape, with corners bending in toward the gap center (Fig. 1d). We attributed this effect to redeposition of sputtered material. As expected, ion beam milling resulted in steep sides of the nano-pyramid frustum (Fig. 1e) [14]. Starting from a gap width of g∼100 nm at the gold top surface, we obtained narrow openings of approximately 15-20 nm at the Au-glass interface (Fig.1 e). We observed that the bottom apertures were occasionally closed, possibly due to Au grains obstructing the beam way [13].

To describe the physics of the device, we performed numerical simulations by means of rigorous coupled-wave analysis (RCWA) method, to solve scattering from periodic structures [16]. Optical data for gold, chromium and borosilicate glass were drawn from the literature [17,18]. We initially simulated the formation of plasmonic bandgaps in arrays of nano-platelets with vertical sidewalls. We simulated the nano-structures with varying slit widths of 10 nm, 15 nm, 20 nm and 30 nm in the H direction, whereas we kept the slit width constant at 25 nm in the V direction. The reflectance and transmittance spectra for periodicity P=640 nm are shown in Fig. 2 a, b for *p* and *s* polarizations, respectively.

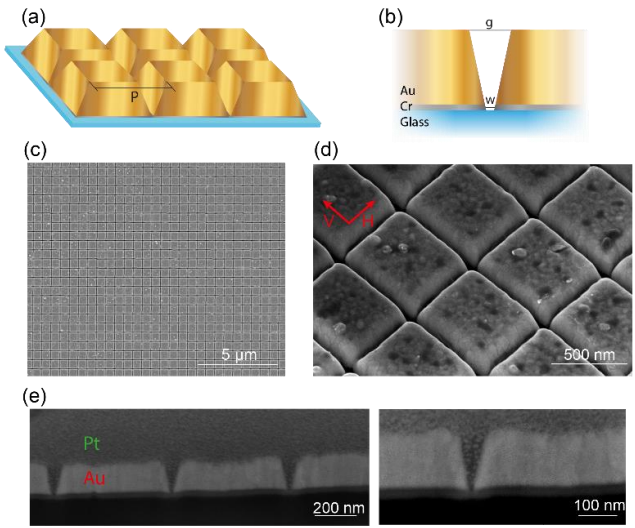

Fig. 1. (a) Sketch of Truncated Nano Pyramid array with periodicity P. (b) Close-up diagram of the tapered groove separating adjacent elements. (c) In-plane SEM micrograph of a TNP array with P=640 nm. (d) Tilted SEM image of TNP array. (e) Cross-sections of the tapered grooves. Platinum was deposited on top of the gold layer to protect layers while milling with FIB.

We observed that the transmittance increases with the gap size up to 30 nm-wide slits (Fig. 2a). At the same time, we confirmed that the different form-factors in H- and V-directions act independently on the bandgap formation, as reported in a previous work [12].

We then modified the simulation to describe the formation of plasmonics bandgaps in TNP arrays. To do this, we simulated gold platelets with steep, inclined sidewalls, forming tapered grooves that are larger at the Au-air interface (Fig. 1b, Fig. 2c, d). To account for fabrication tolerances, we set the bottom aperture to w=15 nm in the H direction and w=25 nm in the V direction. At the same time, we varied the top aperture in the interval 40 nm < g < 70 nm in both H and V directions. The reflectance and transmittance spectra for periodicity P=640 nm are shown in Fig. 2c and 2d for *p* and *s* polarizations, respectively. In the visible region, we observed a lone resonance peak centered on the array periodicity. Interestingly, the groove form-factor enhances the resonance strength when compared with the rectangular slit. This is apparent in the transmittance spectra, where the transmission amplitude roughly doubles for the same aperture width at the gold-glass interface (w=25 nm, Fig. 2b, d). This behavior can be explained by the adiabatic light-funneling properties of tapered grooves, which have been previously employed to enhance extraordinary optical transmission in 1D metal gratings [7,14].

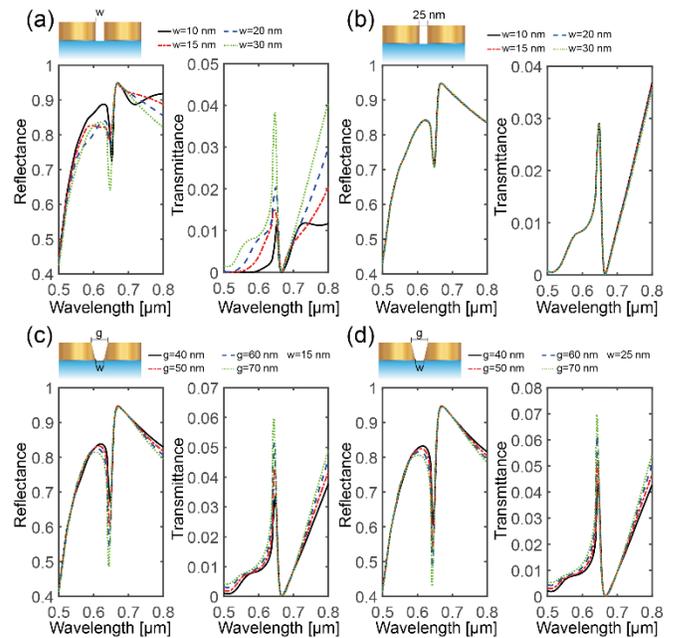

Fig. 2. (a) Reflectance and transmittance spectra for infinite arrays of gold nanoplatelets, with periodicity P=640 nm, for slits of varying width w=10, 15, 20, 30 nm in the H direction and fixed width w=25 nm in the V direction for incident p-polarized light. (b) As in a for incident s-polarized light. (c) Reflectance and transmittance spectra for infinite arrays of TNPs, with periodicity P=640 nm, with bottom aperture fixed at w=15 nm and top aperture g varying as g= 40, 50, 60  70 nm for incident p-polarized light. (d) as in c for fixed bottom aperture w=25 nm for incident s-polarized light.

We acquired experimental reflection and transmission spectra for TNP arrays using a modified upright microscope (Scientica Slicescope). A sketch of the experimental setup is shown in Fig. 3. We illuminated the sample with a collimated broadband lamp, linearly polarized in the range 500-800 nm and spatially filtered with a 100 μm pinhole. We relayed the pinhole image on the sample with a lens (focal length 250 mm), two beam splitters (non-polarizing, 50/50) and an infinity corrected objective (Zeiss 5×, NA=0.16). We collected reflected light with the same objective and transmitted signal with a 20x objective (Olympus, NA=0.50). Spectra analysis was performed with a high-resolution spectrometer equipped with a 300 l/mm grating (Horiba Scientific iHR320, EMCCD camera Horiba Scientific Synapse, 1600 x 1200 pixels).

The experimental data are compared with simulations for two arrays of TNP (array A and array B) illuminated at normal incidence, whose geometrical parameters are reported in Table 1, as extracted from SEM imaging of TNP cross-sections.

**Table 1. Geometrical parameters of simulated TNP arrays**

| Array | Direction | T (nm) | P (nm) | g (nm) | w (nm) |
|---|---|---|---|---|---|
| A | H | 180 | 563 | 86 | 18 |
|   | V | 180 | 572 | 90 | 17 |
| B | H | 180 | 640 | 104 | 17 |
|   | V | 180 | 650 | 110 | 16 |

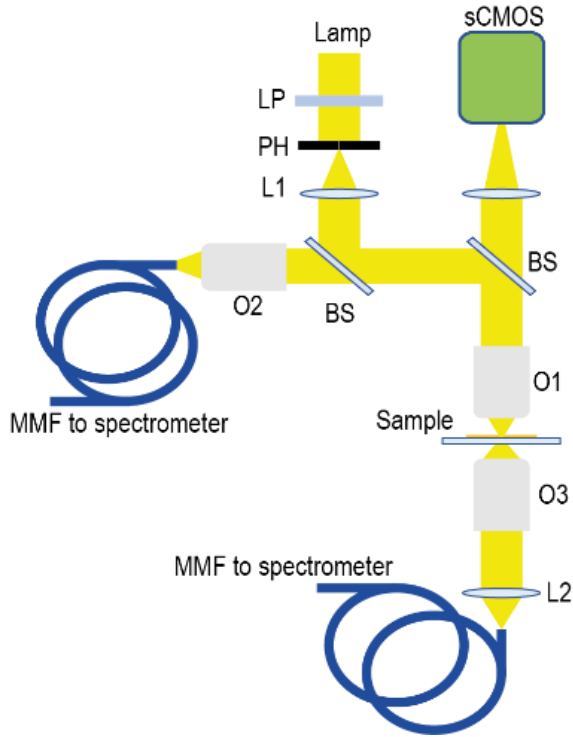

Fig. 3. Sketch of the experimental system: LP, linear polarizer, PH pinhole, L1 lens focal 250 mm, BS 50/50 Beam splitter, O1 5× objective, O2-O3 20× objective, L2 lens focal 40 mm, MMF multimode fiber NA=0.39, sCMOS imaging camera.

The reflectance and transmittance spectra for TNPs arrays A and B are shown in Fig. 4. The experimental reflectance spectra (Fig. 4a, c) confirm the presence of a plasmonic resonance at the array periodicity. The smaller amplitude of the experimental resonance can be ascribed to roughness in the gold layer [19] together with morphological defects arising from FIB milling [13]. In addition, the experimental resonance in array B is split in two dips. A similar effect is less evident in array A because the second dip is mitigated by the rising edge of the bulk gold's reflection. This is possibly linked to oblique light impinging on the sample in the experimental case. At the same time, we experimentally observed a transmission resonance around the array periodicity, as simulated for infinite TNP arrays (Fig. 4b, d).

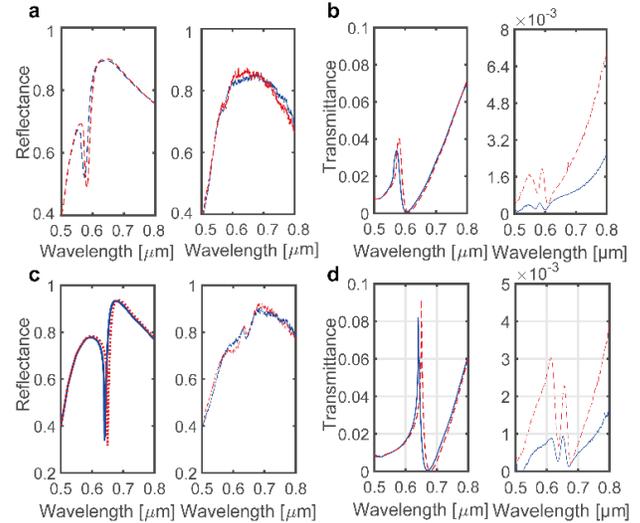

Fig. 4 (a) *Left* Simulated reflectance spectra for the array A with incident light p- (solid blue) and s-(dashed red) polarized light; *right*, measured reflectance spectra for array A with incident p-(solid blue) and s-(dashed red) polarized light. (b) As in *a* for transmittance. (c) As in a for array B; (d) as in b for array B.

The experimental resonance amplitude is smaller than the simulated one because the model considers TNP arrays of infinite size while the experiments are performed on 30 μm×30 μm arrays illuminated by a circular spot ~15 μm in diameter. In addition, transmission resonances are mitigated by surface roughness and morphological defects [13,19].

Despite measured resonances were smaller than the theoretical predictions, the scattering behavior of FIB-milled TNP arrays can be exploited to perform refractive index sensing. Using an upright microscope in dark field (DF) configuration (Nikon Eclipse 2000), we illuminated the sample with oblique light: while the zeroth diffraction order was blocked, the first order was allowed to reach the detector. We acquired DF images with the sample covered in air and in Isopropyl alcohol (IPA). Fig. 5 shows that the scattering spectrum in air (left) is red-shifted with respect to scattering spectrum in IPA (right, RI=1.37). Moreover, it is apparent that only the patterned area produces a diffraction spectrum, as no signal arises from the gold layer around the platelets, apart from scattering from small particles deposited on the sample.

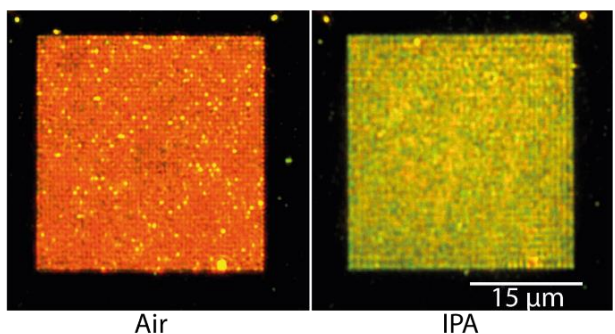

Fig. 5 Dark field images of the TNP array covered in air (left) or IPA (right). The scattered spectrum gets blue-shifted when the Isopropyl alcohol is poured on the TNP array.

In conclusion, we demonstrated the fabrication of 2D arrays of truncated nano-pyramids with sub-wavelength gaps using FIB patterning. We described the effect of FIB milling in the formation of plasmonic bandgaps by modelling light scattering in 2D arrays of gaps with either vertical sidewalls or sharp, tapered edges. This led us to observe plasmonic resonances in the reflectance and transmittance spectra of gold TNP arrays tuned to the visible region. In addition, we successfully detected a RI change in the array surroundings by measuring a spectral shift in the scattered light spectrum. Due to the small sensing area required to detect a RI change, our approach is readily scalable to produce high-density arrays of individually-addressable plasmonic biosensors.

**Funding.** FiP, AB, FeP acknowledge funding from the European Research Council under the European Union's Horizon 2020 research and innovation program (#677683). MP and MDV acknowledge funding from the European Research Council under the European Union's Horizon 2020 research and innovation program (#692643). FeP and MDV acknowledge that project leading to this application has received funding from the European Union's Horizon 2020 research and innovation programme under grant agreement No 828972.

**Acknowledgment**.

**Disclosures**. The authors declare no conflicts of interest.